\documentclass[aps,prl,twocolumn,showpacs,floatfix]{revtex4}
\usepackage{graphicx}
\usepackage{times}
\usepackage{nicefrac}
\usepackage{amsmath}
\usepackage{amsfonts}
\usepackage{amssymb}
\usepackage{amsthm}
\usepackage{epsf}
\usepackage{bm}
\usepackage{bbm}

\usepackage{dcolumn}
\newcolumntype{.}{D{x}{}{-1}}

\newcommand{\be}{\begin{eqnarray}}
\newcommand{\ee}{\end{eqnarray}}
\newcommand{\la}{\langle}
\newcommand{\ra}{\rangle}

\newcommand{\veps}{\varepsilon}

\newcommand{\pr}{\prime}

\newcommand{\rmd}{{\rm d}}

\newcommand{\muN}{\mu_N}
\begin{document}

\title{Screened QED corrections in lithiumlike heavy ions in the
presence of magnetic fields}
\author{A. V. Volotka,$^{1,2}$ D. A. Glazov,$^{1,2}$ V. M. Shabaev,$^{2}$
I. I. Tupitsyn,$^{2}$ and G. Plunien$^{1}$}

\affiliation{
$^1$ Institut f\"ur Theoretische Physik, Technische Universit\"at Dresden,
Mommsenstra{\ss}e 13, D-01062 Dresden, Germany \\
$^2$ Department of Physics, St. Petersburg State University,
Oulianovskaya 1, Petrodvorets, 198504 St. Petersburg, Russia \\
}

\begin{abstract}
A rigorous evaluation of the complete gauge-invariant set
of the screened one-loop QED corrections to the hyperfine
structure and g factor in lithiumlike heavy ions is presented.
The calculations are performed in both Feynman and Coulomb gauges
for the virtual photon mediating the interelectronic interaction.
As a result, the most accurate theoretical predictions for
the specific difference between the hyperfine splitting values
of H- and Li-like Bi ions as well as for the g factor of Li-like
Pb ion are obtained.
\end{abstract}

\pacs{31.30.jf, 31.30.Gs, 31.30.js}

\maketitle
%
%
Investigations of the hyperfine splitting and the g factor in highly charged ions
give an access to a test of bound-state QED in strongest electromagnetic fields
available at present for experimental study.
To date, accurate measurements of the ground-state hyperfine structure
and of the g factor were performed in H-like
$^{209}$Bi, $^{165}$Ho, $^{185}$Re, $^{187}$Re, $^{207}$Pb, $^{203}$Tl, and $^{205}$Tl
\cite{klaft:1994:2425,crespo:1996:826,crespo:1998:879,seelig:1998:4824,beiersdorfer:2001:032506}
and in H-like $^{12}$C \cite{haeffner:2000:5308}
and $^{16}$O \cite{verdu:2004:093002}, respectively.
In particular, the 2002 CODATA value for the electron mass is derived
mainly from the experimental and theoretical g factor values for hydrogenlike
carbon and oxygen with an accuracy 4 times better than that of the 1998 CODATA value.
An extension of such kind of experiments to highly charged
Li-like ions presently being prepared \cite{winters:2007:403,vogel:2008:113}
will provide the possibility to investigate a specific
difference between the corresponding values of H- and Li-like ions,
where the uncertainty due to the nuclear effects can be substantially reduced
\cite{shabaev:2001:3959,glazov:2004:062104,shabaev:2006:253002}.
Achievement of the required theoretical accuracy for the hyperfine structure
and for the g factor in the case of Li-like ions is a very interesting
and demanding challenge for theory.

At present, the theoretical accuracy of the specific difference
of the hyperfine splitting values of H- and Li-like ions
and of the g factor of Li-like heavy ions
is mainly limited by uncertainties
of the screened QED and higher-order
interelectronic-interaction corrections.
In the present Letter we focus on one of the most
difficult correction, namely, the screened QED correction
in the presence of a magnetic field perturbation.
State-of-the-art evaluations of the screened QED correction 
were performed with local screening potentials
\cite{sapirstein:2001:032506,glazov:2006:330,volotka:2008:062507}.
These calculations are based on the well-established technique
developed for the evaluation of the one-loop QED corrections
in the presence of an external potential
\cite{persson:1996:1433}.
However, the employment of a local screening potential does
not allow one to take into account consistently all the contributing diagrams
and to provide a reliable estimation of the uncertainty of the result.
Therefore, a systematic description in the framework of QED
requires the use of perturbation theory. 
This crucial step has been made now and in this Letter we report
on our results of the rigorous evaluation of the complete gauge-invariant
set of the screened one-loop QED corrections.
As the most interesting application of these results towards tests
of the magnetic sector of bound-state QED we present improved
theoretical predictions for the specific difference between
the ground-state hyperfine splitting values of H- and Li-like Bi ions
and for the g factor of Li-like Pb ion.
%

\begin{figure}
\includegraphics[width=0.35\textwidth]{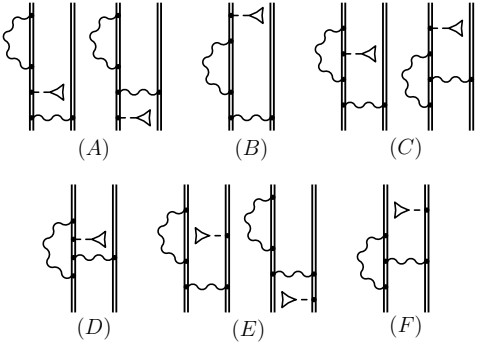}
\caption {Feynman diagrams representing the screened self-energy
correction in the presence of an external potential. The wavy line indicates
the photon propagator and the double line indicates
the electron propagators in the Coulomb field.
The dashed line terminated with the triangle denotes the
interaction with magnetic field.}
\label{scr-se}
\end{figure}

The screened radiative correction in the presence of an external potential
corresponds to the third-order perturbation theory terms.
Nowadays, several approaches are used for derivation of the formal
expressions from the first principles of QED:
the two-time Green-function method \cite{shabaev:2002:119},
the covariant-evolution-operator method \cite{lindgren:2004:161},
and the line profile approach \cite{andreev:2008:135}.
Here, we employ the two-time Green-function method.
To simplify the derivation of formal expressions, we specify
the formalism regarding the closed shell electrons as belonging to
a redefined vacuum.
It implies a modification of the $i0$-prescription in the electron
propagator incorporating the closed-shell electrons.
The corresponding shift of the Fermi-level does not affect
the hyperfine structure and the g factor.
In this way we have to consider all two-loop diagrams
for the valence electron in the presence of magnetic perturbation,
this is 27 nonequivalent diagrams.
These diagrams merge the second-order interelectronic-interaction correction,
the two-loop, and the screened one-loop radiative corrections.
The generic types of the resulting screened self-energy diagrams
are depicted in Fig.~\ref{scr-se}.
The radiative correction to the hyperfine splitting $x_{\rm rad}$
can be written as the sum $x_{\rm rad} = x_{\rm QED} + x_{\rm SQED}$,
where $x_{\rm QED}$ corresponds to the one-electron QED correction,
and $x_{\rm SQED}$ stands for the screened radiative correction.
The latter can be divided into self-energy (SE)
and vacuum-polarization (VP) parts,
$x_{\rm SQED} = x^{\rm SE}_{\rm SQED} + x^{\rm VP}_{\rm SQED}$.
The screened SE correction can be distinguished according
to so-called irreducible (irr) and reducible (red) parts.
It appears as the sum of the following terms: 
\be
\label{atype}
x^{\rm A, irr}_{\rm SQED} = 2 G_a \sum_b \sum_P (-1)^P
\Biggl\{ \sum_n^{\veps_n\neq\veps_a} \la a | \Sigma(\veps_a) | n \ra
\hspace{1cm}
\nonumber\\
\times\frac{\la n b | I(\Delta) | \xi_{Pa} Pb \ra
            + \la n | T_0 | \zeta_{b|PaPb} \ra }{\veps_a-\veps_n}
 + (a \leftrightarrow b)
\Biggr\}\,,
\ee
\be
\label{btype}
x^{\rm B, irr}_{\rm SQED} &=& 2 G_a \sum_b \sum_P (-1)^P
\Bigl\{ \la \xi_a | \Sigma(\veps_a) | \zeta_{b|PaPb} \ra
\nonumber\\
&+&(a \leftrightarrow b)
\Bigr\}\,,
\ee
\be
\label{ctype}
x^{\rm C, irr}_{\rm SQED} = 2 G_a \sum_b \sum_P (-1)^P
\frac{i}{2\pi} \int_{-\infty}^\infty \rmd\omega \sum_{n_1,\,n_2}
\hspace{1cm}
\nonumber\\
\times\Biggl\{
   \frac{\la a n_1 | I(\omega) | n_2 \xi_{Pa} \ra \la n_2 b | I(\Delta) | n_1 Pb \ra}
{(\veps_{Pa}-\omega-u\veps_{n_1})(\veps_a-\omega-u\veps_{n_2})}
\hspace{1.4cm}
\nonumber\\
 +\frac{\la a n_1 | I(\omega) | n_2 \zeta_{b|PaPb} \ra \la n_2 | T_0 | n_1 \ra}
{(\veps_a-\omega-u\veps_{n_1})(\veps_a-\omega-u\veps_{n_2})}
 + (a \leftrightarrow b)
\Biggr\}\,,
\ee
\be
\label{dtype}
x^{\rm D, irr}_{\rm SQED} &=& 2 G_a \sum_b \sum_P (-1)^P
\frac{i}{2\pi} \int_{-\infty}^\infty \rmd\omega \sum_{n_1,\,n_2,\,n_3}
\hspace{1cm}
\nonumber\\
&\times&\Biggl\{
   \frac{\la a n_1 | I(\omega) | n_3 Pa \ra \la n_3 b | I(\Delta) | n_2 Pb \ra}
        {(\veps_{Pa}-\omega-u\veps_{n_1})(\veps_{Pa}-\omega-u\veps_{n_2})}
\nonumber\\
&\times&\frac{\la n_2 | T_0 | n_1 \ra}{(\veps_a-\omega-u\veps_{n_3})}
 + (a \leftrightarrow b)
\Biggr\}\,,
\ee
\be
\label{etype}
x^{\rm E, irr}_{\rm SQED} = 2 G_a \sum_b \sum_P (-1)^P
\Biggl\{ \sum_n^{\veps_n\neq\veps_a} \la a | \Sigma(\veps_a) | n \ra
\hspace{1cm}
\nonumber\\
\times\frac{\la n \xi_b | I(\Delta) | Pa Pb \ra
        + \la n b | I(\Delta) | Pa \xi_{Pb} \ra}
{\veps_a-\veps_n}
 + (a \leftrightarrow b)
\Biggr\}\,,
\ee
\be
\label{ftype}
x^{\rm F, irr}_{\rm SQED} = 2 G_a \sum_b \sum_P (-1)^P
\frac{i}{2\pi} \int_{-\infty}^\infty \rmd\omega \sum_{n_1,\,n_2}
\hspace{1.2cm}
\nonumber\\
\times\Biggl\{
   \frac{\la a n_1 | I(\omega) | n_2 Pa \ra \la n_2 \xi_b | I(\Delta) | n_1 Pb \ra}
{(\veps_{Pa}-\omega-u\veps_{n_1})(\veps_a-\omega-u\veps_{n_2})}
 + (a \leftrightarrow b)
\Biggr\}\,,
\ee
\be
\label{gtype}
x^{\rm G, red}_{\rm SQED} = 2 \, G_a \sum_b \sum_P (-1)^P
\Biggl\{
   \la a | \Sigma(\veps_a)
\Biggl[
   | \zeta^\pr_{b|PaPb} \ra
\hspace{0.6cm}
\nonumber\\
\times \la a | T_0 | a \ra
 + | \xi^\pr_a \ra \la a b | I(\Delta) | Pa Pb \ra
 + | a \ra \la \zeta^\pr_{b|PaPb} | T_0 | a \ra
\nonumber\\
 + | a \ra \Bigl( \la a | T_0 | a \ra - \la b | T_0 | b \ra \Bigr)
    \frac{\la a b | I^{\prime\prime}(\Delta) | Pa Pb \ra}{2}
    \delta_{bPa}
\hspace{0.65cm}
\nonumber\\
 + | \xi_a \ra \la a b | I^\prime(\Delta) | Pa Pb \ra
 + | a \ra \la a b | I^\prime(\Delta) | \xi_{Pa} Pb \ra
\hspace{1.1cm}
\nonumber\\
 - \sum_n^{\veps_n\neq\veps_a} 
    \frac{\la n b | I^\prime(\Delta) | Pa Pb \ra \la b | T_0 | b \ra}
    {\veps_a-\veps_n}\, | n \ra
\Biggr]
 + (a \leftrightarrow b)
\Biggr\}\,,
\hspace{0.2cm}
\ee
\be
\label{htype}
x^{\rm H, red}_{\rm SQED} = G_a \sum_b \sum_P (-1)^P
\Biggl\{
 \frac{i}{2\pi} \int_{-\infty}^\infty \rmd\omega \sum_{n_1,\,n_2}
\hspace{1.1cm}
\nonumber\\
\times \Biggl[
   \frac{\la a n_1 | I(\omega) | n_2 a \ra \la n_2 | T_0 | n_1 \ra
         \la a b | I^\prime(\Delta) | Pa Pb \ra}
        {(\veps_a-\omega-u\veps_{n_1})(\veps_a-\omega-u\veps_{n_2})}
\hspace{1.05cm}
\nonumber\\
 + 2 \frac{\la a n_1 | I(\omega) | n_2 Pa \ra \la n_2 b | I^\prime(\Delta) | n_1 Pb \ra
         \la a | T_0 | a \ra}
        {(\veps_{Pa}-\omega-u\veps_{n_1})(\veps_a-\omega-u\veps_{n_2})}
   \Biggr]
\hspace{0.85cm}
\nonumber\\
 +  2\la a | \Sigma^\pr(\veps_a)
\Biggl[
   | \xi_a \ra \la a b | I(\Delta) | Pa Pb \ra
 + | \zeta_{b|PaPb} \ra \la a | T_0 | a \ra
\nonumber\\
 + | a \ra \Bigl( \la a | T_0 | a \ra - \frac{\la b | T_0 | b \ra}{2} \Bigr)
   \la a b | I^\prime(\Delta) | Pa Pb \ra
\hspace{1.5cm}
\nonumber\\
 + | a \ra \Bigl( \la \zeta_{b|PaPb} | T_0 | a \ra + \la \zeta_{a|PbPa} | T_0 | b \ra \Bigr)
\Biggr]
 + (a \leftrightarrow b)
\Biggr\}\,,
\ee
\be
\label{itype}
x^{\rm I, red}_{\rm SQED} = - G_a \sum_b \sum_P (-1)^P
\Biggl\{
\frac{i}{2\pi} \int_{-\infty}^\infty \rmd\omega \sum_{n_1,\,n_2}
\hspace{1.1cm}
\nonumber\\
\times \Biggl[
   \frac{\la a n_1 | I(\omega) | n_2 a \ra \la a b | I(\Delta) | Pa Pb \ra}
        {(\veps_a-\omega-u\veps_{n_1})(\veps_a-\omega-u\veps_{n_2})}
   \Biggl(\frac{\la n_2 | T_0 | n_1 \ra}{\veps_a-\omega-u\veps_{n_1}}
\hspace{0.5cm}
\nonumber\\
        +\frac{\la n_2 | T_0 | n_1 \ra}{\veps_a-\omega-u\veps_{n_2}}\Biggr)
 + \frac{\la a n_1 | I(\omega) | n_2 Pa \ra \la n_2 b | I(\Delta) | n_1 Pb \ra}
        {(\veps_{Pa}-\omega-u\veps_{n_1})(\veps_a-\omega-u\veps_{n_2})}
\nonumber\\
\times\Biggl(\frac{\la Pa | T_0 | Pa \ra}{\veps_{Pa}-\omega-u\veps_{n_1}}
        +\frac{\la a | T_0 | a \ra}{\veps_a-\omega-u\veps_{n_2}}\Biggr)
   \Biggl]
\hspace{2.25cm}
\nonumber\\
 - \la a | \Sigma^{\prime\prime}(\veps_a) | a \ra \la a | T_0 | a \ra
   \la a b | I(\Delta) | Pa Pb \ra
 + (a \leftrightarrow b)
\Biggr\}\,.
\hspace{0.5cm}
\ee
Here, $a$ and $b$ denotes the valence and core electron states, respectively,
the sum over $b$ runs over all closed-shell states,
$P$ is the permutation operator, giving rise to the sign $(-1)^P$ of the permutation,
and the notation $(a \leftrightarrow b)$
stands for the contribution with interchanged labels $a$ and $b$.
The SE operator $\Sigma(\veps)$, the interelectronic-interaction
operator $I(\omega)$, and their derivatives (indicated by primes)
are defined similar as in Ref.~\cite{shabaev:2002:119}, $u=1-i0$ preserves
the proper treatment of poles of the electron propagators.
The energy difference $\Delta$ is defined as $\Delta = \veps_a - \veps_{Pa}$
and, accordingly, $\Delta = \veps_b - \veps_{Pb}$ in terms $(a \leftrightarrow b)$.
$T_0$ is the electronic part of the hyperfine-interaction operator
and $G_a$ is the multiplicative factor depending on the quantum numbers
of the valence electron (see for details Ref.~\cite{volotka:2008:062507}).
The wavefunctions $|\xi\ra$, $|\zeta\ra$, and $|\xi^\pr\ra$, $|\zeta^\pr\ra$
are defined as follows
\be
\label{xi-oxi_a-1}
| \xi_a \ra = \sum_n^{\veps_n\neq\veps_a}
\frac{| n \ra \la n | T_0 | a \ra}{\veps_a-\veps_n}\,,
\ee
\be
\label{zeta_a-1}
| \zeta_{b|PaPb} \ra = \sum_n^{\veps_n\neq\veps_a}
\frac{| n \ra \la n b | I(\Delta) | Pa Pb \ra}{\veps_a-\veps_n}\,,
\ee
and
$| \xi^\pr_a \ra = \frac{\partial}{\partial\veps_a} | \xi_a \ra$,
$| \zeta^\pr_{b|PaPb} \ra = \frac{\partial}{\partial\veps_a} | \zeta_{b|PaPb} \ra$.

The expressions compiled in Eqs.~(\ref{atype})-(\ref{ctype}), and (\ref{etype})-(\ref{htype})
contain ultraviolet divergences.
We separate out the divergent zero- and one-potential terms in
Eqs.~(\ref{atype}),~(\ref{btype}),~(\ref{etype}),~(\ref{gtype})
and zero-potential terms in
Eqs.~(\ref{ctype}),~(\ref{ftype}),~(\ref{htype})
and evaluate these terms in the momentum space,
where the divergences can be removed analytically
(see, e.g., Ref.~\cite{yerokhin:1999:800}).
The remaining many-potential terms are ultraviolet finite
and calculated in coordinate space.
The infrared divergences which occur in the terms of the
Eqs.~(\ref{ctype}),~(\ref{dtype}),~(\ref{ftype}),~(\ref{htype}),~(\ref{itype})
are regularized by introducing a nonzero photon mass and
canceled analytically.

The numerical evaluation is based on
employing the dual-kinetic-balance finite basis
set method \cite{shabaev:2004:130405} with basis functions
constructed from B-splines \cite{sapirstein:1996:5213}.
The Fermi model for the nuclear charge density and the sphere model for
the magnetic moment distribution have been employed.
In what follows we present our result for the case of Li-like
Bi utilizing the corresponding values for the nuclear properties:
$\la r^2 \ra^{1/2} = 5.5211$ fm \cite{angeli:2004:185},
$I^\pi=9/2-$, and $\mu=4.1106(2) \muN$ \cite{stone:2005:75}.
The calculations have been performed in Feynman and Coulomb gauges
for the photon propagator describing the electron-electron interaction,
thus providing an accurate check of the numerical procedure.
The obtained results for the screened SE correction
for the hyperfine splitting of the Li-like Bi
are presented in Table~\ref{tab:gauges} in both gauges, respectively.
\begin{table}
\caption{Individual contributions to the screened SE
correction $x^{\rm SE}_{\rm SQED}$ for the ground-state hyperfine structure
of the Li-like $^{209}$Bi$^{80+}$.}
\label{tab:gauges}
\tabcolsep3pt
\begin{tabular}{lrrlrr}                                                    \hline
Contr. &    Feynman  &    Coulomb  & Contr. &    Feynman  &    Coulomb  \\ \hline
A, irr &    0.001544 &    0.001555 & F, irr & $-$0.000174 & $-$0.000172 \\
B, irr & $-$0.000380 & $-$0.000398 & G, red & $-$0.001298 & $-$0.001307 \\
C, irr &    0.001928 &    0.001952 & H, red &    0.000331 &    0.000331 \\
D, irr & $-$0.000936 & $-$0.000945 & I, red &    0.000066 &    0.000066 \\
E, irr &    0.000028 &    0.000028 &        &             &             \\ \hline
       &             &             & Total  &    0.001109 &    0.001109 \\ \hline
\end{tabular}
\end{table}
Finally, we have calculated the screened SE correction within local
screening potentials: Kohn-Sham $0.0012$ and core-Hartree $0.0013$.
The results are in reasonable agreement with the rigorous evaluation
$x^{\rm SE}_{\rm SQED} = 0.00111$.

We have also calculated the screened VP correction in the presence
of a magnetic field employing the Uehling approximation for the VP loop.
The results have been checked utilizing the Feynman and Coulomb gauges
for the photon propagator mediating the interelectronic interaction.
The electric-loop part of the screened Wichmann-Kroll (WK) contribution
has been calculated by means of the approximate formulas for the WK
potential from Ref.~\cite{fainshtein:1990:559}.
As concerns the screened WK magnetic-loop part
we have employed the hydrogenic $2s$ value from Ref.~\cite{artemyev:2001:062504},
assuming that it enters with the same screening ratio
as the Uehling terms.
Accordingly, our value for the screened VP correction is
$x^{\rm VP}_{\rm SQED} = -0.00054(2)$.
Finally, the total value for the screened QED correction to the
ground-state hyperfine structure in Li-like Bi
results as $x_{\rm SQED} = 0.00057(2)$.

Probing the influence of QED effects on the hyperfine splitting
of highly charged ions is impeded by the uncertainty of the
nuclear magnetization distribution correction [the Bohr-Weisskopf (BW) effect].
In this context, it was proposed to consider a specific difference of the
ground state hyperfine splitting in H-like and Li-like ions
\cite{shabaev:2001:3959}:
$\Delta^\prime E = \Delta E^{(2s)} - \xi \Delta E^{(1s)}$,
where $\Delta E^{(1s)}$ and $\Delta E^{(2s)}$ are the hyperfine
splittings of H- and Li-like ions, respectively,
the parameter $\xi$ is chosen to cancel the BW correction.
In this specific difference the nuclear corrections almost
vanish completely.
We have recalculated the energy shift $\Delta^\prime E$
employing the most accurate result obtained for the screened QED correction.
The interelectronic-interaction corrections
have been evaluated to first-order in $1/Z$
within the QED perturbation theory and to higher-orders within
the large-scale configuration-interaction Dirac-Fock-Sturm method.
Extracting numerically the contribution of the BW effect
in different terms we found that the cancellation
appears with $\xi$ chosen to be $\xi = 0.16886$ for the case of Bi.
In Table~\ref{tab:diff} we present obtained result for the specific
difference between the hyperfine structure values of H- and Li-like Bi ions.
The Dirac value incorporates also the finite nuclear size correction.
\begin{table}
\caption{Individual contributions to the specific
difference $\Delta^\prime E$ for $^{209}$Bi in meV.}
\label{tab:diff}
\begin{tabular}{lrrr}                                                                                   \hline
                      &  $\Delta E^{(2s)}$\;\; &  $\xi \Delta E^{(1s)}$  & $\Delta^\prime E$\;\;\; \\ \hline
Dirac value                           &   844.829\phantom{(4)} &  876.638 & $-$31.809\phantom{(4)} \\
Interel. inter.,$\sim 1/Z$            & $-$29.995\phantom{(4)} &          & $-$29.995\phantom{(4)} \\
Interel. inter.,$\sim 1/Z^2$ and h.o. &     0.25(4)\;\;        &          &     0.25(4)\;\;        \\
QED                                   &  $-$5.052\phantom{(4)} & $-$5.088 &     0.036\phantom{(4)} \\
Screened QED                          &     0.194(6)           &          &     0.194(6)           \\
Total                                 &                        &          & $-$61.32(4)\;\;        \\ \hline
\end{tabular}
\end{table}
The nuclear-polarization correction to the $1s$ hyperfine splitting
calculated in Ref.~\cite{nefiodov:2003:35} yields
$\xi \Delta E^{(1s)}_{\rm NP} = 0.009$ meV.
However, since all nuclear corrections have the similar
scaling dependence upon the principal quantum numbers,
we expect the strong cancellation between $1s$ and $2s$
nuclear-polarization corrections in the specific difference.
The same is valid for the second-order one-electron QED contributions.
Comparing with the results for the specific difference
$\Delta^\prime E$ presented in Ref.~\cite{shabaev:2001:3959}
we have increased the accuracy for the screened QED part
and performed more elaborate calculations
for the higher-order interelectronic-interaction correction.
Further rigorous evaluation of the higher-order electron-electron
interaction corrections will provide a test of bound-state QED
at strongest electric and magnetic fields.

Similar calculations have been performed for the g factor of Li-like heavy ions.
Here, we present our results for the case of $^{208}$Pb$^{79+}$
with the following value for the nuclear charge radius
$\la r^2 \ra^{1/2} = 5.5010$ fm \cite{angeli:2004:185}.
The rigorous evaluation of the screened SE correction gives
$\Delta{\rm g}^{\rm SE}_{\rm SQED} = -3.1(1) \times 10^{-6}$.
The previous value obtained with local screening potentials
was $-3.5(1.2) \times 10^{-6}$ \cite{glazov:2006:330}.
Thus, the uncertainty of the screened SE correction has been
reduced by an order of magnitude.
The screened VP contribution has been calculated within
the Uehling approximation.
As to the WK part, we have employed the approximate formulas
for the electric-loop potential \cite{fainshtein:1990:559},
while the magnetic-loop value has been taken from Ref.~\cite{lee:2005:052501},
assuming the same screening ratio as for the Uehling term.
Accordingly, we have obtained $\Delta{\rm g}^{\rm VP}_{\rm SQED} = 1.5 \times 10^{-6}$.
In Table~\ref{tab:g} we have updated the value for the g factor
of Li-like $^{208}$Pb$^{79+}$ previously reported in Ref.~\cite{glazov:2006:330}
employing the result obtained for the screened QED correction.
\begin{table}
\caption{Individual contributions to the ground-state g factor
of Li-like $^{208}$Pb$^{79+}$.}
\label{tab:g}
\tabcolsep5pt                           
\begin{tabular}{ll}                                             \hline
Dirac value (point nucleus) & \phantom{$-$} 1.932\,002\,904    \\
Finite nuclear size         & \phantom{$-$} 0.000\,078\,58(13) \\
Interel. inter.             & \phantom{$-$} 0.002\,140\,7(27)  \\
QED, $\sim \alpha$          & \phantom{$-$} 0.002\,411\,7(1)   \\
QED, $\sim \alpha^2$        &          $-$  0.000\,003\,6(5)   \\
Screened QED                &          $-$  0.000\,001\,6(1)   \\
Nuclear recoil              & \phantom{$-$} 0.000\,000\,25(35) \\
Nuclear polarization        &          $-$  0.000\,000\,04(2)  \\
Total                       & \phantom{$-$} 1.936\,628\,9(28)  \\ \hline
\end{tabular}
\end{table}
Further extensions of these calculations to the g factor of B-like heavy
ions may serve for an independent determination of the fine structure constant
from QED at strong fields \cite{shabaev:2006:253002}.
%

In summary, we have rigorously calculated the screened QED correction
to the hyperfine splitting and g factor of heavy Li-like ions.
We have increased the theoretical accuracy for the specific difference
between the hyperfine splitting values of H- and Li-like bismuth
as well as for the g factor of Li-like lead.
The rigorous calculation of the higher-order interelectronic-interaction
correction will be the next step towards the unprecedented
accuracy for the stringent test of the bound-state QED
in the presence of magnetic fields.
%

The authors acknowledge financial support from DFG, GSI,
RFBR (Grant No. 07-02-00126a), and Ministry of Education and Science of
Russian Federation (Program for Development of Scientific Potential of
High School, Grant No. 2.1.1/1136).
The work of D.A.G. was also supported by
the grant of President of Russian Federation
(Grant No. MK-3957.2008.2) and by the FAIR - Russia Research Center.
%
%

%
\end{document}